\documentclass[sigconf]{acmart}
\AtBeginDocument{%
  \providecommand\BibTeX{{%
    \normalfont B\kern-0.5em{\scshape i\kern-0.25em b}\kern-0.8em\TeX}}}

\setcopyright{acmcopyright}
\copyrightyear{2023}
\acmYear{2023}
\acmDOI{XXXXXXX.XXXXXXX}

\acmConference['GEC 2023]{ACM Gender Equality in Computing}{June 27,
  2023}{Athens University of Economics and Business, Athens, Greece}
\begin{document}


\title{A Blockchain-based Electronic Voting System: EtherVote}

\author{Achilleas Spanos}
\affiliation{%
  \institution{Department of Informatics and Computer Engineering,\\
  University of West Attica}
  \streetaddress{Ag. Spyridonos Str., Egaleo}
  \city{Athens}
  \country{Greece}
  \postcode{12243}}
\email{cs161048@uniwa.gr}

\author{Ioanna Kantzavelou}
\authornotemark[1]
\orcid{1234-5678-9012}
\affiliation{%
  \institution{Department of Informatics and Computer Engineering,\\
  University of West Attica}
  \streetaddress{Ag. Spyridonos Str., Egaleo}
  \city{Athens}
  \country{Greece}
  \postcode{12243}}
\email{ikantz@uniwa.gr}



\begin{abstract}

The development of an electronic voting system that would replace traditional election procedures is a research topic of great interest for many years. Blockchain technology could provide some guarantees and fulfill strong requirements for electronic voting platforms, such as transparency, immutability, and confidentiality. From time to time research is conducted to address problems in voting systems. Many research works attempt to implement secure and reliable voting systems, which address known security, anonymity, and fraud issues that might threaten such systems.
 
This paper presents a proposal of a secure electronic voting system, the EtherVote, using the Ethereum Blockchain network that focuses deeply on the field of identification of eligible citizens. The proposed system will be entirely based on Blockchain without any central authority servers or databases, thus improving security, privacy, and election cost. Limitations, problems, and solutions are discussed, in order to make the proposed electronic voting system ideal and ready to use for national elections.

\end{abstract}

\ccsdesc[300]{Security and privacy~Security protocols}

\keywords{Blockchain, Voting, Ethereum, Smart Contract, Metamask}


\maketitle


\section{Introduction}

There is a number of supportive reasons to replace in-person voting procedures or other special voting facilities, such as absentee voting, voting in a foreign country, early voting, or proxy voting, with electronic voting (e-voting). Through e-voting, equal voting rights could be provided to citizens facing access problems, such as people with disabilities, and long-distance coverage. Nevertheless maintaining and storing the votes in a database, which would be managed by an organization, incorporates the risks of running into over-authority and manipulated details, limiting fundamental fairness, privacy, anonymity, and transparency in the voting process. Central authorities could delete or modify votes. Even if the authority is trustworthy, an attacker could gain access to the database and modify or change votes and personal data. At the same time, old-fashioned paper voting is very complicated to verify and audit for a citizen who has no control over the voting system. Blockchain is a new and promising technology that could be used to address these problems towards e-voting solutions. The full potential of this technology is not yet implemented \cite{Kalogiratos2022}.

With the use of blockchain technology, the main requirements and difficulties of a voting procedure, vote integrity, and security could be addressed. Following the blockchain structure and functioning, adding the vote to a new block creates a reference to the previous one. Thus, an immutable chain is created, in which once a piece of information is added - in this case a vote - it cannot be modified without destroying block relations. Thus, the technologies offered by the blockchain combined with the proper use of an encryption algorithm to protect votes, aiming at the anonymity of the votes, but also with a secure citizen identification system could permanently change the way of voting for each country. At the same time, any citizen will be able to verify his vote, as opposed to the traditional way of voting, by getting the transaction hash value. A hash value verifies the transaction with blockchain, which is casting a vote.

A thorough survey \cite{9812616} that provides a complete comparison between the very recent Blockchain-based methods adopted by electronic voting systems, establishes the state of the art and exposes the achievements of such efforts. Using these outcomes as a point of start, we propose and implement a system that will be an improvement on the already existing electronic voting systems.

In this research work, we propose an electronic voting system, the EtherVote, based entirely on the Ethereum blockchain using smart contracts. Since user identification is a major problem in such systems and proposals, especially when no database or classic server is used, we will focus on adding or combining some authentication factors to validate eligible citizens with the right to vote, before and during the voting procedure. On platforms that are expected to serve any kind of social group, of all ages, ease of use is a critical factor. The Ethereum network is public and can be easily accessed by Metamask, - a self-custodial wallet to safely access blockchain applications - whose addresses will be matched with citizens. That makes the user authentication, the voting procedure, and generally any interaction with the blockchain very easy to use.

However, the choice of a public blockchain network brings about the need for encryption of citizens' personal data as well as votes. This encryption is implemented by the selection of appropriate cryptographic algorithms.

\section{Proposed system}

In the proposed system EtherVote, the only service provided by any authority is to record the list of eligible citizens on the blockchain. The procedure of identifying citizens with the right to vote, the voting procedure, and the storage of votes will be based exclusively on the storage of data and the call of functions in the smart contract. Every transaction, whether for identification or for voting is public and can be easily traced, while keeping all personal data private. Systems that use the blockchain as a database, inherit the immutability and therefore modification or deletion of information is impossible. Each voting citizen will be assigned a Metamask address - with a necessary number of ethers, during the entire election procedure. The electoral procedure is divided into four phases, as described in the sequel.

The \textbf{first phase} consists of writing and storing the smart contract on the Ethereum blockchain. Initially, Metamask addresses are created, which will be considered as 'trusted' and will belong to the electoral authorities. Via these addresses, the smart contract will be created, and the results of the elections will be received, but also the sensitive personal data of citizens will be stored in the blockchain, aiming at using these personal data during authentication.

In the \textbf{second phase}, every eligible citizen is asked to register on the platform. In order to register, he must either attend or contact the authorities. The authorities must be connected to one of the 'trusted' addresses (presented in \textbf{phase 1} ), and after creating a new Metamask account, with which the citizen will be assigned and identified, is registered on the platform by linking the newly created address with the citizen's personal data, such as ID number, first name, last name, and phone number. Although the variables that will store the information are private, since the Ethereum network is public, anyone with a copy of the blockchain will be able to retrieve this private information. To address this risk, personal data are combined, then encrypted with the cryptographic hash algorithm SHA256, and matched as a key-value pair with the address assigned to each citizen. Upon completion of the second phase, each citizen receives the password for the Metamask account with which they have been matched.

The \textbf{third phase} is the process of identifying voters on election day. Once they enter the voting platform, they must log in to their Metamask account by entering the password they received when registered, thus creating the first identification parameter. Next, as a second identification parameter, they will have to connect to the platform, entering their personal information, with which the specific address has been assigned. Similar to stage 2, the data is combined and encrypted with SHA256, followed by a check to match the transaction's sending address with the hash. If the second authentication stage is successfully completed, a unique code (OTP) will be generated. This code is stored in the Blockchain, at the time of its creation, and it is matched with the address corresponding to the current voter, as a key-value pair, just as the match was made with the personal data. Saving the time the OTP was created is to have a window of time to use it. Finally, each voter receives this unique code via SMS to his/her mobile phone that is registered.

At last, the \textbf{fourth phase} is the voting procedure. Each voter having performed the identification, and having received the unique code on his/her mobile, is invited to vote, choosing from a list of candidates, which is also stored in the Blockchain. When the candidate is selected, the unique code received during identification is requested in order to accept the vote. In the case that the OTP is incorrect, or the time limit has expired, he must receive a new OTP, performing the second stage of identification. In the opposite case, the transaction is done, the number of votes of each candidate is increased, and after the vote is registered, the state of the address of the voter is changed, so that it is locked and it is not possible to add more than one vote. If the current Metamask address has registered a vote on the Blockchain, in that case, at the second stage of identification, a unique password will not be generated and sent, and there is no possibility of submitting a second vote.

\section{Discussion and Conclusions}

The system EtherVote has been implemented and achieved to fulfill the most important requirements, immutability and anonymity in voting.  Some weaknesses need to be addressed in the future, to make EtherVote ideal and ready for use in several electoral procedures, even at the level of national elections.

The first weakness that needs to be resolved concerns the sending of the one-time password via SMS. It is not possible to send SMS, but also, in general, to call APIs from the smart contract. Also, in our system implementation, we do not take particular measures regarding the storage of votes, as the names and the number of candidates may differ according to the needs of each voting process. Therefore, for each voting case, appropriate methods for secure vote storage could be used, which would make the vote storage safe and untraceable.

Oracles could be a solution to these problems, so we could call APIs from the smart contract, but also store the votes of each candidate. This is a third-party service solution that could intervene between smart contracts and external systems. They can provide smart contracts with access to external data and services, including the ability to make HTTP requests. Finally, the use of a suitable cryptographic algorithm compliant with voting needs could also be ideal for the safe storage of votes.

The use of Blockchain technology could guarantee the integrity of the votes and could make e-voting applications secure, easy to use, and low cost, while satisfying at the same time all the principles of elections. Identification and authentication, but also the requirement of anonymity in voting are sufficiently covered in the proposed research work. With the addition of sending an SMS and the appropriate encryption of personal data, enhances the security of this system that could be ideal for national elections. As the technologies offered by the Blockchain and smart contracts improve day by day, the existence of an even more improved e-voting application is a matter of time.


\bibliographystyle{ACM-Reference-Format}
\bibliography{Spanos_arxiv}


\end{document}